\documentclass[sigconf]{acmart}

\AtBeginDocument{%
  }

\copyrightyear{2026}
\acmYear{2026}
\setcopyright{cc}
\setcctype{by}
\acmConference[KDD '26]{Proceedings of the 32nd ACM SIGKDD Conference on Knowledge Discovery and Data Mining V.2}{August 09--13, 2026}{Jeju Island, Republic of Korea}
\acmBooktitle{Proceedings of the 32nd ACM SIGKDD Conference on Knowledge Discovery and Data Mining V.2 (KDD '26), August 09--13, 2026, Jeju Island, Republic of Korea}
\acmDOI{10.1145/3770855.3818444}
\acmISBN{979-8-4007-2259-2/2026/08}

\usepackage{graphicx}
\usepackage{booktabs}
\usepackage{multirow}
\usepackage{arydshln}
\usepackage{xcolor}
\usepackage{multicol}

\usepackage[ruled,vlined,linesnumbered]{algorithm2e}

\begin{document}

\title{MatchLM2Lite: A Scalable MLLM-to-Lite Framework for Reproduced Content Identification}

\author{Xiaotian Fan}
\authornote{Equal contribution.}
\affiliation{%
  \institution{Tiktok}
  \city{Singapore}
  \country{Singapore}
}
\email{xiaotian.fan@tiktok.com}

\author{Hiok Hian Ong}
\authornotemark[1]
\affiliation{%
  \institution{Tiktok}
  \city{Singapore}
  \country{Singapore}
}
\email{hiok.ong@tiktok.com}

\author{David Yuchen Wang}
\authornotemark[1]
\affiliation{%
  \institution{Tiktok}
  \city{Singapore}
  \country{Singapore}
}
\email{david.w@tiktok.com}

\author{Zirui Zhu}
\affiliation{%
  \institution{Tiktok}
  \city{Singapore}
  \country{Singapore}
}
\affiliation{%
  \institution{National University of Singapore}
  \department{School of Computing}
  \city{Singapore}
  \country{Singapore}
}
\email{zirui.zhu@tiktok.com}

\author{Kanchan Sarkar}
\authornote{Project lead.}
\authornote{Corresponding author.}
\affiliation{%
  \institution{Tiktok}
  \department{Content Intelligence}
  \city{San Jose}
  \state{CA}
  \country{USA}
}
\email{kanchan.sarkar@tiktok.com}

\author{Kun Xu}
\affiliation{%
  \institution{Tiktok}
  \city{San Jose}
  \state{CA}
  \country{USA}
}
\email{daniel.chen28@tiktok.com}

\renewcommand{\shortauthors}{Xiaotian Fan et al.}

\begin{abstract}
Content moderation is critical for online video platforms to ensure content safety, protect creators, and sustain positive user experiences. Beyond filtering harmful content, platforms must guarantee content authenticity at scale so that users are exposed to diverse, original videos rather than low-value reproductions. We present \textbf{MatchLM2Lite}, a real-time, production-grade reproduced content identification (RCI) system that leverages the powerful understanding of a multimodal large language model (MLLM) distilled into a small and fast-inference model. Our system jointly models video, audio, and text signals, operating on pairs of videos to produce fine-grained reproduction scores.
The system comprises two modules, \textbf{MatchLM} and \textbf{MatchLite}, and a two-stage training recipe. First, our high-capacity MLLM, MatchLM, serves as a teacher model to define the upper bound of RCI performance. Its capabilities are then distilled into a compact student model, MatchLite. This design allows MatchLite to deliver low-latency, high-throughput inference on video pairs while preserving much of MatchLM’s accuracy, making it suitable for integration into real-time recommendation systems.
MatchLM achieves an F1-score improvement of \textbf{+8.57} compared to our previous production model. After knowledge distillation, MatchLite retains a \textbf{+6.55} gain in F1-score while reducing computational cost by \textbf{35$\times$}. Deployed at scale, MatchLM2Lite enables efficient, pairwise multimodal RCI, stably serving online traffic at high queries per second (QPS) with an end-to-end latency below 30 seconds. This system has reduced the reproduced video view rate on our platform by \textbf{2.5\%} without degrading user engagement, demonstrating its effectiveness in a large-scale production environment.
\end{abstract}



\maketitle

\section{Introduction}
Content moderation has become a crucial concern for ensuring video quality and user experiences on short video platforms such as Instagram Reels, Pinterest, TikTok, and Kuaishou, helping to provide healthy and creative content while reducing harmful and low-quality content \cite{vlm_as_policy, shortvideo_service}. 
\begin{figure*}[h]
  \centering
  \includegraphics[width=\linewidth]{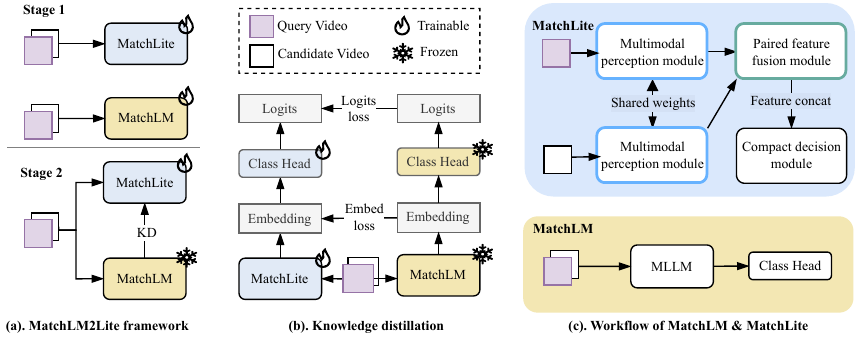}
  \caption{Overall design of MatchLM2Lite (a) MatchLM2Lite Framework involves two stages of training: Stage 1 - Teacher \& student trained separately with GT labels. Stage 2 - Freeze teacher; train student with supervised + distillation losses; (b) Knowledge Distillation. Student learns from frozen MLLM via cosine embedding loss and KL divergence on logits, along with standard cross-entropy loss; (c) Workflow of MatchLM \& MatchLite.}
  \label{fig:two_stage_training}
\end{figure*}

Content authenticity protection is a key part of content moderation, aiming to reduce the spread of copied or minimally modified videos. These efforts protect the rights of content creators, enhance content diversity, improve recommendation quality, and contribute to a more sustainable and safe ecosystem. 

The core technical challenge is reproduced content identification (RCI) which refers to the identification of variants modified by trimming, filtering, or editing between query and candidate videos. It requires joint modeling of paired videos across multiple modalities to capture detailed similarities and overall semantic information. Existing methods mainly rely on embedding-based similarity using visual encoders \cite{vcsl, transvcl, SVD}, ignoring key text and audio signals which users engage with when watching short videos. 

To address these limitations, we propose a \textbf{MatchLM2Lite} framework consisting of a high-capacity multimodal large language model \textbf{MatchLM}, for setting a strong reference point for model performance, along with a lightweight multimodal model \textbf{MatchLite}, for real-time serving. Both models jointly encode visual, audio and text, enabling stronger cross-modal interactions. 

We train this framework in a two-stage approach. In the first stage, \textbf{MatchLM} and \textbf{MatchLite} are trained in a supervised manner using labeled data. In the second stage, \textbf{MatchLM} is frozen and acts as a "teacher" to distill its knowledge into \textbf{MatchLite} \cite{mm_distill}. This distillation process enables \textbf{MatchLite} to inherit the capabilities of \textbf{MatchLM} with strong cross-modal alignment and rich semantic representation, while maintaining its compact architecture suitable for large-scale production deployment.

We validate the proposed \textbf{MatchLM2Lite} framework through extensive offline experiments and online A/B testing on a large-scale, real-world short video platform. It shows strong effectiveness in production content governance workflows, with full-scale deployment supporting a throughput of over 3k requests per second.

Our key contributions are as follows:
\begin{itemize}
\item We introduce a unified multimodal framework for reproduced content identification, formulating and modeling it as a video pair matching problem across three modalities: visual, audio, and text. We apply joint encoding and alignment of multimodal signals to enable robust reproduced content identification.
\item We design a \textbf{MatchLM2Lite} system that consists of a MLLM-based teacher model (\textbf{MatchLM}) with a lightweight student model (\textbf{MatchLite}) for efficient real-time deployment. \textbf{MatchLM} learns rich cross-modal representations for classification, while \textbf{MatchLite} maintains semantic alignment and efficiency through knowledge distillation.
\item We deploy our approach at scale on our short video platform and achieve consistent improvements in both offline and online settings. Compared to our prior production model, \textbf{MatchLM} achieves an F1 improvement of +8.57; \textbf{MatchLite} achieves an F1 of +6.55 after knowledge distillation. Online A/B tests show a 2.5\% reduction in reproduced content views, demonstrating production effectiveness.
\end{itemize}
\section{Related Works}
\label{sec: Related Works}
\subsection{Content Moderation}
Content moderation is crucial for shielding users from harmful content and protecting creators from plagiarism \cite{video_similarity, policy_as_prompt}. Human-based content moderation demands significant human labor \cite{community_moderation}, can cause emotional distress through exposure to large amounts of toxic/harmful content, and is susceptible to biases affecting fairness and consistency \cite{moderation_bias}. Model-driven moderation can alleviate both economic and psychological costs of human moderation while enhancing the safety and quality of online content. Previous works have employed Neural Networks for toxic-content detection \cite{perspective_ai, harmful_content_detection, transformer_offensive_content}, as well as localization modules for video-copy detection \cite{vcsl, transvcl, rtr}. Online systems typically engage large-scale retrieval systems such as \cite{faiss}, which are used to find candidate matches for a given query. This generates paired samples which then need to be evaluated for content duplication \cite{vcsl}, reproduction, or other policy-dependent violations \cite{yew2026dynamic}. 
 
\subsection{Multi-Modal Interaction with LLMs}
The development of multi-modal large language models (MLLMs)\cite{llava_one_vision, qwen25vl, internvid25, video_llava} has employed various techniques to enhance video comprehension in conjunction with language. LLaVA-One-Vision \cite{llava_one_vision} is among one of the top performing open-source MLLMs which leverages the pre-trained Qwen-2 \cite{qwen2} language backbone and SigLIP \cite{siglip} vision encoder. Existing small audio models such as Whisper \cite{whisper} and related works \cite{viola, voxtlm, spectron} focus narrowly on specific audio domains like human speech or natural sounds \cite{pengi}. Efforts have aimed to extend MLLMs to handle diverse audio inputs, by injecting audio information into pretrained backbones. QwenAudio \cite{qwenaudio} integrates diverse audio signals through an early-fusion approach to encode audio embeddings into a large language model. Video-LLaMA \cite{video_llama} utilizes separate branches for different modalities and leverages existing pretrained embeddings from ImageBind \cite{imagebind} to learn visual-audio-language correspondence. Qwen-Omni \cite{Qwen3-Omni} similarly integrates audio, visual and text modalities and trains an end-to-end MLLM to achieve holistic perception capabilities. These works demonstrate success in leveraging pretrained large vision-language model weights to improve performance in additional modalities. 

\subsection{Industrial Applications}

Recent works have explored the potential of MLLMs for tasks such as recommendation and content moderation. For example, NoteLLM2 \cite{notellm2} proposes joint end-to-end finetuning of existing LLMs and vision encoders to create strong multi-modal embeddings of visual and text content, which is utilized in their content recommendation systems. Similarly, QARM \cite{qarm} extends this by incorporating audio information into their MLLM, further enhancing recommendation capabilities. Kuaishou also demonstrated that vision-language models can serve as effective online content moderators  \cite{vlm_as_policy}, highlighting the versatility of MLLMs in diverse business applications. More recently, Qwen3-VL-Reranker \cite{qwen3reranker} has emerged as a strong open-source multimodal reranking framework, providing a state-of-the-art baseline for query-document relevance estimation across mixed modalities. As MLLMs rapidly evolve in their capacity to accurately model human preferences \cite{zhu2026camel,christiano2017deep, dai2024safe}, this has also catalyzed growing interest in leveraging their capabilities for advanced recommendation and content moderation.

\subsection{Knowledge Distillation}
One disadvantage of using MLLMs is their large size and high latency, which is resource-intensive. Knowledge distillation \cite{knowledge_distillation} can distill the knowledge from large, cumbersome models down to smaller-sized models more suited to production environments. Previous works have shown that large LLM/MLLMs can serve as effective teachers to teach both smaller-scale LLM/MLLMs \cite{multi_teacher_kd, llava_mod, llava_kd}, or other small models, achieving performance surpassing the student model itself. The predicted target distributions from the teacher can be directly used as a loss objective for the predicted distributions for the student \cite{distilledbert}.

\section{Methodology}

The main challenge of reproduced content identification (RCI) lies in detecting a variety of modifications between paired videos while avoiding penalizing the non-copied videos (false positives). RCI is an internal task which seeks to identify videos that offer little to no transformative value over existing content already on the platform, fully considering visual, audio, and textual signals. To the best of our knowledge, no similar public benchmarks exist. The complex policy is not dictated by fixed quantitative thresholds; instead, trained annotators determine whether  sufficient original value remains, guided by internal annotation guidelines and curated examples for consistency. Full policy details are not disclosed to prevent adversarial exploitation, and we provide representative case studies illustrating these policy-dependent decisions in Appendix~\ref{sec:Case Studies}. Existing solutions for reproduced video detection typically utilize a visual encoder and focus only on matching visual similarities \cite{vcsl, transvcl, rtr}. Some previous in-house solutions for this task propose multi-tower retrieval frameworks that separate each modality into standalone recall pipelines with late-stage fusion modules. However, this design leads to greater system complexity, higher serving latency, and also weakens cross-modal alignment.

\subsection{MatchLM2Lite Training Framework}
To address these limitations, we propose the \textbf{MatchLM2Lite} framework to balance accuracy and efficiency. In this section, we introduce the architectural design of \textbf{MatchLM} and \textbf{MatchLite} and the end-to-end two stage training recipe for the framework, as illustrated in Figure~\ref{fig:two_stage_training} (a).
\begin{figure}[h]
  \centering
  \includegraphics[width=0.5\textwidth]{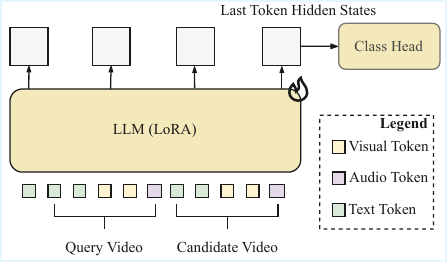}
  \caption{MatchLM architecture design. Query and candidate video embeddings are interleaved and fed into the LLM. The final hidden state of the last token serves as the video representation, followed by a classification head for prediction.}
  \label{fig:img1_mllm_overall}
\end{figure}
In first stage training, both \textbf{MatchLM} and \textbf{MatchLite} are trained for paired video matching across modalities and  are independently finetuned with supervised ground-truth labels. \textbf{MatchLM} adopts an early-fusion strategy by transforming visual frames, audio, and accompanying text into token sequences and uses multimodal projectors for modality alignment. The token sequences are then processed by an LLM backbone fine-tuned via LoRA \cite{hu2022lora}. \textbf{MatchLite} uses efficient multimodal encoders to extract and fuse modality-specific features, as shown in Figure~\ref{fig:two_stage_training} (c).

In the second stage, knowledge distillation is performed to transfer the learned multimodal semantic representations of \textbf{MatchLM} to \textbf{MatchLite}. Specifically, \textbf{MatchLM} parameters are frozen, while \textbf{MatchLite} is further trained using a combination of distillation loss and supervised classification loss, leveraging the pretrained checkpoints from stage one, as shown in Figure ~\ref{fig:two_stage_training} (b). More details are provided in section ~\ref{sec:method_kd}.

\subsection{MatchLM Architecture Design}
We build \textbf{MatchLM} upon LLaVA-One-Vision (0.5B) (LLaVA-OV) \cite{llava_one_vision}: the architecture incorporates SigLIP as the visual encoder and leverages Qwen2 as the LLM backbone. We extend the model to simultaneously process audio \cite{video_llama, qwenaudio}. 

We begin with separate encoding strategies for each modality to obtain their respective token representations. For the visual and text modalities, we adopt the default LLaVA-One-Vision \cite{llava_one_vision} pipeline, where video frames are sampled at fixed intervals, encoded via SigLIP-So400m-Patch14-384 \cite{siglip}, and injected into the LLM through the original vision-to-language projector.

To incorporate audio, inspired by QwenAudio \cite{qwenaudio}, we use the Whisper-small \cite{whisper} encoder to extract high-level representations from raw audio. The input audio is sampled at 16kHz and transformed into 80-dimensional log-mel features, from which we extract 1500 audio tokens per video. A special \texttt{<audio>} token is inserted to denote the position of the audio embeddings in the input sequence. Additionally, a learnable audio saliency weighting layer is used to further aggregate the sequence of audio tokens into a single audio token. Instead of using Q-Former \cite{video_llama,blip2,llava} as a modality bridge like in Video-LLaMA \cite{video_llama}, we follow the same design in LLaVA-One-Vision to use  a lightweight MLP projector \cite{llava, llava_one_vision} to map audio tokens to the same embedding dimension as text and visual input embeddings. The final input to the LLM consists of interleaved tokens from all modalities, beginning with task-specific prompts.
To enable pairwise video comparison, we extend the standard MLLM input format to support two-video inputs—a query video and a candidate video—within a single forward pass, as shown in Figure ~\ref{fig:img1_mllm_overall}. The input sequence is constructed as:
\[
\texttt{Prompt} + [\texttt{Video}_1] + [\texttt{Audio}_1] + [\texttt{Video}_2] + [\texttt{Audio}_2].
\]

We utilize the MLLM model as a paired video representation extractor \cite{gme,mm_embed} rather than as a next-token prediction generative model. This design enables direct use of rich semantic embeddings for discriminative classification and helps effectively distill knowledge into a lightweight \textbf{MatchLite} model. Specifically, we use the final hidden state corresponding to the last token as the paired video representation \cite{coefvq}. 
On top of the LLM output, we append a lightweight classification head. The representation is passed through a shared projection layer followed by task-specific output layers. 

This structure supports various content reproduction classification tasks while maintaining low inference cost. We use cross-entropy loss for each task and aggregate them as the final multi-task classification objective:
\begin{equation}
\mathcal{L}_{\text{class}} = \sum_{t=1}^{T} \mathcal{L}_{\text{CE}}^{(t)}
\label{eq:ce_loss}
\end{equation}

In practice, this consists of the main RCI task prediction layer alongside other auxiliary task prediction layers for sublabels (e.g. subtitles) which are discarded at inference time.

\subsection{MatchLite Architecture Design}
Next, we focus on \textbf{MatchLite}, the lightweight student model designed for efficient deployment. As shown in Figure~\ref{fig:two_stage_training} (c), \textbf{MatchLite} consists of three main modules: Multimodal Perception, Paired Feature Fusion, and a Compact Decision Module.

\begin{figure}[t]
  \centering
  \includegraphics[width=0.5\textwidth]{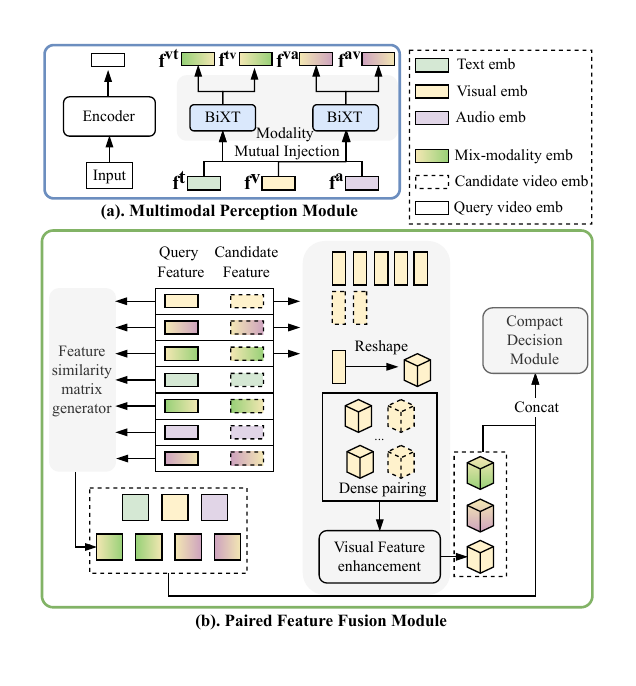}
  \vspace{-40pt} 
  \caption{Main modules of MatchLite. (a). Multimodal Perception Module: Frozen encoders extract modality-specific features. BiXT applies bidirectional cross-attention between modality pairs; (b). Paired Feature Fusion Module: Computes cosine similarity between query-candidate feature pairs. Visual features are further fused via lightweight CNNs. All outputs are concatenated for final decision.}
  \label{fig:img4_matchLite}
  \vspace{-10pt}
\end{figure}

The Multimodal Perception Module utilizes pretrained, frozen encoders—Swin Transformer~\cite{swint} for vision, Sentence-BERT~\cite{wang2020minilm} for text, and Whisper-small~\cite{whisper} for audio. Modality-specific features $f$ for each video $V_x$ ($x \in \{q, c\}$), are extracted yielding $f^v_x$ (visual), $f^t_x$ (text), and $f^a_x$ (audio), where $q$ denotes query and $c$ denotes candidate. To further enhance visual representations, we employ a modality mutual injection mechanism based on bidirectional cross-attention (BiXT)~\cite{bixt}, which produces four mixed-modality embeddings for each video:
\begin{align}
  [f^{vt}_x,\, f^{tv}_x] &= \mathrm{BiXT}^{(v \leftrightarrow t)}(f^v_x,\, f^t_x), \notag \\
  [f^{va}_x,\, f^{av}_x] &= \mathrm{BiXT}^{(v \leftrightarrow a)}(f^v_x,\, f^a_x),
\end{align}
where $f^{vt}_x$, $f^{tv}_x$, $f^{va}_x$, and $f^{av}_x$ denote the mixed-modality embeddings for video $V_x$. The process is shown in Figure ~\ref{fig:img4_matchLite} (a).

The Paired Feature Fusion Module jointly processes the features of the query and candidate videos. For each modality $m \in {v, va, vt, a, av, t, tv}$, we collect the corresponding pair of features $[f^m_q, f^m_c]$. For visual-related modalities ($m \in {v, va, vt}$), a dense pairwise tensor $$D^m_{i,j} = \mathrm{Concat}(f^m_{q,i}, f^m_{c,j})$$ is constructed and passed through several CNN layers to obtain enhanced pairwise features $E^m$. In parallel, cosine similarity matrices $$S^m_{i,j} = \mathrm{cosine}(f^m_{q,i}, f^m_{c,j})$$ are computed for all modalities. All enhanced feature maps $E^{m}$ and similarity matrices $S^{m}$ are concatenated as $\mathcal{F}$ for downstream processing. The overall procedure is summarized in Algorithm~\ref{alg:matchlite_algo}.

\begin{algorithm}[t]
\caption{MatchLite reproduced content identification}
\label{alg:matchlite_algo}
\KwIn{Query video $V_q$, Candidate video $V_c$}
\KwOut{Match score $s$}

\BlankLine
\textbf{// Multimodal Perception}

Initialize $\mathcal{O} = \emptyset$\;
\For{each $x \in \{q, c\}$}{
    $f^v_x \gets \mathrm{Encoder}^{(v)}(V^v_x)$\;
    $f^t_x \gets \mathrm{Encoder}^{(t)}(V^t_x)$\;
    $f^a_x \gets \mathrm{Encoder}^{(a)}(V^a_x)$\;
    $[f^{vt}_x,\, f^{tv}_x] \gets \mathrm{BiXT}^{(v \leftrightarrow t)}(f^v_x,\, f^t_x)$\;
    $[f^{va}_x,\, f^{av}_x] \gets \mathrm{BiXT}^{(v \leftrightarrow a)}(f^v_x,\, f^a_x)$\;
}

\For{each modality $m \in \{v, va, vt, a, av, t, tv\}$}{
    $o^m = [f^m_q, f^m_c]$\;
    Append $o^m$ to $\mathcal{O}$\;
}

\BlankLine
\textbf{// Paired Feature Fusion}

Initialize $\mathcal{E} = \emptyset$, $\mathcal{S} = \emptyset$\;
\For{each $o^m \in \mathcal{O}$}{
    $[f^m_q,\, f^m_c] = o^m$\;
    \If{$m \in \{v, va, vt\}$}{
        $D^m_{i,j} = \mathrm{Concat}\left(f^m_{q,i},\ f^m_{c,j}\right),\ i \in [1,M],\, j \in [1,N]$\;
        $E^m = \mathrm{ResBlockConv}(D^m)$\;
        Append $E^m$ to $\mathcal{E}$\;
    }
    $S^m_{i,j} = \mathrm{CosineMap}(f^m_{q,i},\, f^m_{c,j})$\;
    Append $S^m$ to $\mathcal{S}$\;
}
$\mathcal{F} = \mathrm{Concat}(\mathcal{E},\, \mathcal{S})$\;

\BlankLine
\textbf{// Compact Decision}

$z = \mathrm{ResNet34}\left( \mathcal{F} \right )$\;
$s = \mathrm{MultiTaskHead}(z)$\;
\Return $s$\;
\end{algorithm}

All fused features $\mathcal{F}$ are passed to a lightweight ResNet-34-based Compact Decision Module, which produces the final multimodal representation for classification. A multi-task classification head further supports multiple downstream tasks, shown in Figure ~\ref{fig:img4_matchLite} (b).

\subsection{Two-Stage Training Recipe}
\subsubsection{Stage 1: Supervised training}
We train the \textbf{MatchLM} and \textbf{MatchLite} independently using standard supervised training for paired video classification using the same training data and task objectives, as shown in Figure~\ref{fig:two_stage_training} (a). The supervision loss is formulated as the sum of cross-entropy losses across all downstream classification tasks, shown as Equation ~\ref{eq:ce_loss}.

\subsubsection{Stage 2: Knowledge distillation}
\label{sec:method_kd}
Although \textbf{MatchLM} achieves strong performance, deploying it at scale will incur significant GPU overhead. Therefore, to improve efficiency during deployment, we adopt a knowledge distillation strategy \textbf{MatchLM2Lite} to transfer its capabilities to \textbf{MatchLite}, as illustrated in Figure~\ref{fig:two_stage_training} (b). We apply both embedding distillation and logit distillation.

For embedding-level distillation, we extract the final hidden state of the last input token from \textbf{MatchLM} as a unified paired video-level representation as well as the last embedding state from \textbf{MatchLite}. A learnable projection head is applied to the MLLM outputs to transform it into the dimension of the student embeddings. We then minimize the cosine distance between the student representation $\mathbf{z}_s$ and the teacher representation $\mathbf{z}_t$:

\begin{equation}
\mathcal{L}_{\text{emb}} = 1 - \cos(\mathbf{z}_s, \mathbf{z}_t) = 1 - \frac{\mathbf{z}_s \cdot \mathbf{z}_t}{|\mathbf{z}_s| |\mathbf{z}_t|}
\end{equation}

For logit distillation, we apply a Kullback–Leibler (KL) divergence loss to align the predictive distributions between \textbf{MatchLM} and \textbf{MatchLite}, on the softmax of their predicted logits.
\begin{equation}
\mathcal{L}_{\text{logits}} = \mathrm{KL}\left( \sigma\left( \frac{\mathbf{p}_t}{T} \right) \Big| \sigma\left( \frac{\mathbf{p}_s}{T} \right) \right)
\end{equation}

Here, $\sigma(\cdot)$ denotes the softmax function.

Our two-part distillation loss combines both embedding-level and logit-level training objectives and is given by:

\begin{equation}
\mathcal{L}_{\text{distill}} = \mathcal{L}_{\text{emb}} + \mathcal{L}_{\text{logits}}
\end{equation}

Finally, the overall training objective linearly combines the distillation loss with the task-specific classification loss:

\begin{equation}
\mathcal{L}_{\text{total}} = \lambda \cdot \mathcal{L}_{\text{distill}} + \mathcal{L}_{\text{class}}
\end{equation}

We set the weighting coefficient $\lambda = 1.5$ to strengthen the influence of distillation versus classification during training. 
Our embedding distillation is formulated with a cosine objective rather than $\ell_1$ or $\ell_2$ regression. This enforces only directional consistency between $\mathbf{z}_s$ and $\mathbf{z}_t$, leaving the student free to choose a representation suited to its own architecture. This is desirable when the teacher and student are heterogeneous: \textbf{MatchLM} produces a global semantic embedding through autoregressive token interactions, whereas \textbf{MatchLite} explicitly parameterizes pairwise relations through dense visual similarity maps and task-specific inductive biases. The learnable projection head $g(\cdot)$ aligns these two representation spaces while preserving the student's structural priors. The complementary KL term on classification logits provides a second, low-dimensional signal anchored at the task output. Together, these two losses form a representation-level signal (cosine, structural) and a decision-level signal (KL, task-aligned) that jointly transfer the teacher's knowledge despite architectural differences. This formulation is in line with recent cross-architecture KD work such as VL2Lite \cite{vl2lite}, which uses pairwise similarity-relation alignment to transfer knowledge from large vision-language models to lightweight networks. Empirically, our ablations in Section~\ref{subsec:Effect of Knowledge Distillation} confirm this: logit KL alone yields R@P80 = 65.34\%, and adding the cosine embedding term raises it to 66.92\% (+1.58 \%), validating that embedding-level alignment captures complementary knowledge beyond logit matching alone.

\section{Experiments}
\label{sec: Experiments}

\subsection{Dataset}
\label{sec:Dataset details: Reproduced Content Identification (RCI) Dataset}
We construct a multimodal, video-paired RCI dataset using videos from our platform. For this, we leverage an in-house recall pipeline based on visual features. We first sample a group of query videos, and recall the top 1 candidate video based on visual similarity for each query to form a query/candidate pair. These pairs are sent for human-labeling to obtain their final RCI label. In total, we obtain 0.8 million video pairs with 4.77\% being reproduced pairs. 

To ensure comprehensive coverage, we curate our dataset by generating query videos by 1) random sampling daily published videos, which represent the common types of normal/reproduced content, and 2) sampling from high engagement videos, which are daily videos that have gone viral or pseudo-viral and thus garnered higher popularity, which represent specialized cases of reproduced content. This ensures that the dataset distribution accounts for both the large majority of normal/reproduced content types as well as those at greater risk of being reproduced due to their high user viewership. 

We treat RCI as a binary classification task, and focus on the following metrics: F1-score, Average Precision (AP), and recall at precision 80\% (R@P80). R@P80 is used as an important metric to ascertain the expected proportion of reproduced content that can be recalled after the model is deployed online. Our end goal is to identify reproduced content with high precision and apply suppression strategies on copied video content while also ensuring minimal overkill (false positive) cases for the full platform traffic.

\subsection{Preliminary Exploration}

We perform preliminary explorations to test the zero-shot capabilities of existing open/close sourced models on our RCI task. We first investigate powerful MLLMs including GPT-4o \cite{gpt4o}, Qwen2.5VL \cite{qwen25vl}, LLaVA-OV \cite{llava_one_vision}, Qwen3-VL-Reranker \cite{qwen3reranker} (both 2B and 8B), as well as the open-source TransVCL model \cite{transvcl}, which is one of the best models on video-copy detection. Despite these MLLMs having strong general multimodal capabilities and TransVCL being pretrained on large-scale video copy-localization data, all the models yielded relatively poor performance, as shown in Table ~\ref{tab:pretrained_mlllm_comparison}. Notably, even the Qwen3-VL-Reranker family achieves R@P80 of 0.0, indicating that off-the-shelf multimodal reranking capability is insufficient for our policy-grounded RCI task. These results motivate the need for supervised finetuning on our in-house RCI training set.

\begin{table}[t]
  \centering
  \caption{Zero-shot evaluations on our RCI test set using existing models. For TransVCL, \underline{cls}: uses the cls-token with a linear layer as a classifier and \underline{localization}: uses the detected bounding box directly from TransVCL to assess whether the video contains a duplicated segment.}
  \label{tab:pretrained_mlllm_comparison}
  \vspace{-2pt} 
  \small
  \setlength{\tabcolsep}{3pt}
  \resizebox{\columnwidth}{!}{%
  \begin{tabular}{@{}lccrr@{}}
    \toprule
      Model Name       & Open sourced & Modalities  & \textbf{F1} & \textbf{R@P80}\\
      \midrule
      TransVCL (cls) & Y & V & 14.84 & 0.0\\
      TransVCL (localization) & Y & V & 15.17 & 0.0\\
      GPT-4o & N & V+T & 22.69 & 0.0\\
      Qwen2.5VL 3B & Y & V+T & 8.6 & 0.0\\
      LLaVA-OV 0.5B & Y & V+T & 12.2 & 0.0\\
      Qwen3-VL-Reranker 2B & Y & V+T & 55.15 & 0.0\\
      Qwen3-VL-Reranker 8B & Y & V+T & 55.85 & 0.0\\
      + SFT on in-house data & Y & V+A+T & 77.25 & --\\
      \bottomrule
  \end{tabular}%
  }
  \vspace{2pt}\\
  {\footnotesize Note: For GPT-4o we use model version gpt-4o-2024-11-20}
  \vspace{-10pt}
\end{table}

\subsection{Main Experiment Results}
\label{sec:Main Experiment Results}

The results of our model training on our RCI dataset are presented in Table ~\ref{tab:table1_ablations_lm_lite}. As a baseline, we train a visual-only version of \textbf{MatchLite}, without the use of the Multimodal Perception Module. We find that incorporating audio and text modalities significantly improves upon the visual only baseline model. 
We further perform data scaling experiments to dive into the sample efficiency of the models and show that in-house pretraining also boosts model performance.
Finally, by integrating Knowledge Distillation, we bridge the gap in performance between the teacher \textbf{MatchLM} and the student \textbf{MatchLite}. With all techniques applied, \textbf{MatchLite} achieves AP=82.45$\pm$0.09 and F1=77.10$\pm$0.08, while \textbf{MatchLM} achieves AP=86.21$\pm$0.08 and F1=79.12$\pm$0.30, with standard deviations estimated from three experimental runs with additional random seeds. The teacher model, \textbf{MatchLM} outperforms the \textbf{MatchLite} in AP (+3.76) and F1 (+2.02). Our training hyperparameter details can be found in Appendix ~\ref{sec:Training Details}.

\begin{table*}[t]
  \centering
  \caption{Experiment results on input modality, training data size, and knowledge distillation. We compare the results for MatchLite and MatchLM across the integration of different modalities (visual/audio/text) and perform experiments using different data scales. Finally, we show the improvements in performance when MatchLite is distilled from MatchLM. 
The baseline model configuration is \underline{underlined}, the best MatchLite configuration is \textbf{bolded} and '+' refers to the addition of in-house task pretraining dataset
}
\label{tab:table1_ablations_lm_lite}
\vspace{-4pt} 
  \begin{tabular}{p{3cm}lccc@{\hspace{0.3em}}cc@{\hspace{0.3em}}cc@{\hspace{0.3em}}c}
    \toprule
    \textbf{Setting} & \textbf{Model} & \textbf{Modality} & \textbf{Data (\%)} & \textbf{AP} & & \textbf{F1} & & \textbf{R@P80} & \\
    \midrule
    
    \multirow{6}{*}{\textbf{Modality}} 
      &  \underline{Baseline}     & V             & \multirow{4}{*}{100} & \underline{74.70}         & & \underline{70.55} & & \underline{43.40} &              \\
      &  \multirow{3}{*}{MatchLite}    & V + T         &  & 75.53 & (+0.83) & 71.49 & (+0.94) & 43.58 & (+0.18)     \\
      &       & V + A         &  & 78.09 & (+3.39) & 73.81 & (+3.26) & 54.61 & (+11.21)     \\
      &       & V + A + T     &  & 78.38 & (+3.68) & 74.11 & (+3.56) & 55.52 & (+12.12)     \\
      \cmidrule(lr){2-10} 
      & \multirow{2}{*}{MatchLM}        & V + T         & \multirow{2}{*}{100}  & 81.00 & & 76.26 & & 58.16 &                \\
      &         & V + A + T     &  & 84.99 & (+3.99)  & 78.53 & (+2.27) & 71.21 & (+13.05)     \\
      
    \midrule

    \multirow{8}{*}{\textbf{Data Size}} 
      & \multirow{3}{*}{MatchLite}      & \multirow{3}{*}{V + A + T}     & 33  & 77.08        & & 72.84 & & 49.21 &              \\
      &       &     & 66  & 78.14 & (+1.06) & 73.68 & (+0.84) & 50.25 & (+1.04)             \\
      &       &    & 100 & 78.38 & (+1.30) & 74.11 & (+1.27) & 55.52 & (+6.31)      \\

    \cmidrule(lr){2-10}

      & MatchLite+ &  V + A + T  & 100+  & 80.59 & (+3.51) & 74.48 &(+1.64) & 57.21 & (+8.00) \\
    \cmidrule(lr){2-10}

      & \multirow{3}{*}{MatchLM}        &  \multirow{3}{*}{V + A + T}    & 33  & 82.63         & & 77.53 & & 65.46 &              \\
      &         &      & 66  & 83.15 & (+0.52) & 77.63 & (+0.10) & 65.96 & (+0.50)     \\
      &         &      & 100 & 84.99 & (+2.36) & 78.53 & (+1.00) & 71.21 & (+5.75)     \\
    \cmidrule(lr){2-10} 
      & MatchLM+ &   V + A + T  & 100+  & 86.21 &(+3.58) & 79.12 & (+1.59) & 75.92 & (+10.46) \\
      
    \midrule

    \multirow{4}{*}{\textbf{Distillation}} 
      & MatchLite+ (SwinT, w/o KD)    & \multirow{4}{*}{V + A + T} & \multirow{4}{*}{100+} & 80.59 & & 74.48 & & 57.21 &  \\
      & MatchLite+ (SwinL, w/o KD)    & &  & 82.73 & (+2.14) & 75.49 & (+1.01) & 69.81 & (+12.60) \\
      & \textbf{MatchLite+ (SwinT, w KD)}*   & &  & \textbf{82.45} & (+1.86) & \textbf{77.10} & (+2.62) & \textbf{66.92} & (+9.71) \\
      & MatchLM+              &  &  & 86.21 & & 79.12 & & 75.92 &  \\
      
    \bottomrule
\end{tabular}
\end{table*}

\begin{table*}[ht]
  \centering
  \caption{\textbf{MatchLM} Ablation Studies: We compare the effects of using Last Token Classification with Next Token Prediction (NTP), Early audio fusion with Late audio fusion, Dynamic vs Static Frame Allocation and the difference in performance when using Qwen2.5VL 3B backbone vs LLaVA-OV 0.5B.}
  \vspace{-4pt}
    \label{tab:mllm_ablation_studies}
    \begin{tabular}{lccccc}
      \toprule
      \textbf{\textbf{MatchLM} Model}        & \textbf{Objective} & \textbf{Audio Fusion} & \textbf{Frame Allocation} &
      \textbf{AP} & \textbf{F1}\\
      \midrule
      Qwen2.5VL 3B & Last Token Cls & Early & Dynamic & 83.38 & 77.45 \\
      LLaVA-OV 0.5B & NTP & Early & Dynamic & 83.13 & 77.81 \\
      LLaVA-OV 0.5B & Last Token Cls & Late & Dynamic & 83.51 & 77.65 \\
      LLaVA-OV 0.5B & Last Token Cls & Early & Static & 84.83 & 77.86 \\
      LLaVA-OV 0.5B & Last Token Cls & Early & Dynamic & \textbf{84.99} & \textbf{78.53} \\
      \bottomrule
    \end{tabular}
    
\end{table*}

\begin{table}
  \vspace{-8pt}
  \centering
  \caption{Ablation study on Knowledge Distillation on the best performing MatchLite+ (with the addition of in-house task pretraining dataset): The effect of varying $\mathcal{L}_{\text{logits}}$ and $\mathcal{L}_{\text{emb}}$ shows that performance improves and then plateaus at $\mathcal{L}_{\text{logits}}$=1.5 and $\mathcal{L}_{\text{emb}}$=1.5.}
  \label{tab:model-comparison-lightweight-kd-mllm}
  \vspace{-14pt} 
  \begin{tabular}{lcclll@{\hspace{-0.1em}}p{0.7cm}}
    \toprule
    Model&$\mathcal{L}_{\text{logits}}$&$\mathcal{L}_{\text{emb}}$&AP&F1&R@P80&\\
    \midrule
    \textbf{MatchLite+}& -- & -- & 80.59 & 74.48 & 57.21 &\\
    \textbf{MatchLite+} & 1.0 & 1.0 & 81.95 & 76.75 & 64.26\\
    \textbf{MatchLite+} & 1.5 & 1.5 &  \textbf{82.45} & 77.10 & \textbf{66.92} & (+9.71) \\
    \textbf{MatchLite+} & 2.0 & 2.0 & 82.37 & 77.29 & 65.70\\
  \bottomrule
\multicolumn{5}{l}{\footnotesize + add In-house Task Pretraining} \\
\end{tabular}

  \vspace{-20pt} 
\end{table}

\begin{figure}[h]
  \centering
  \includegraphics[width=\linewidth]{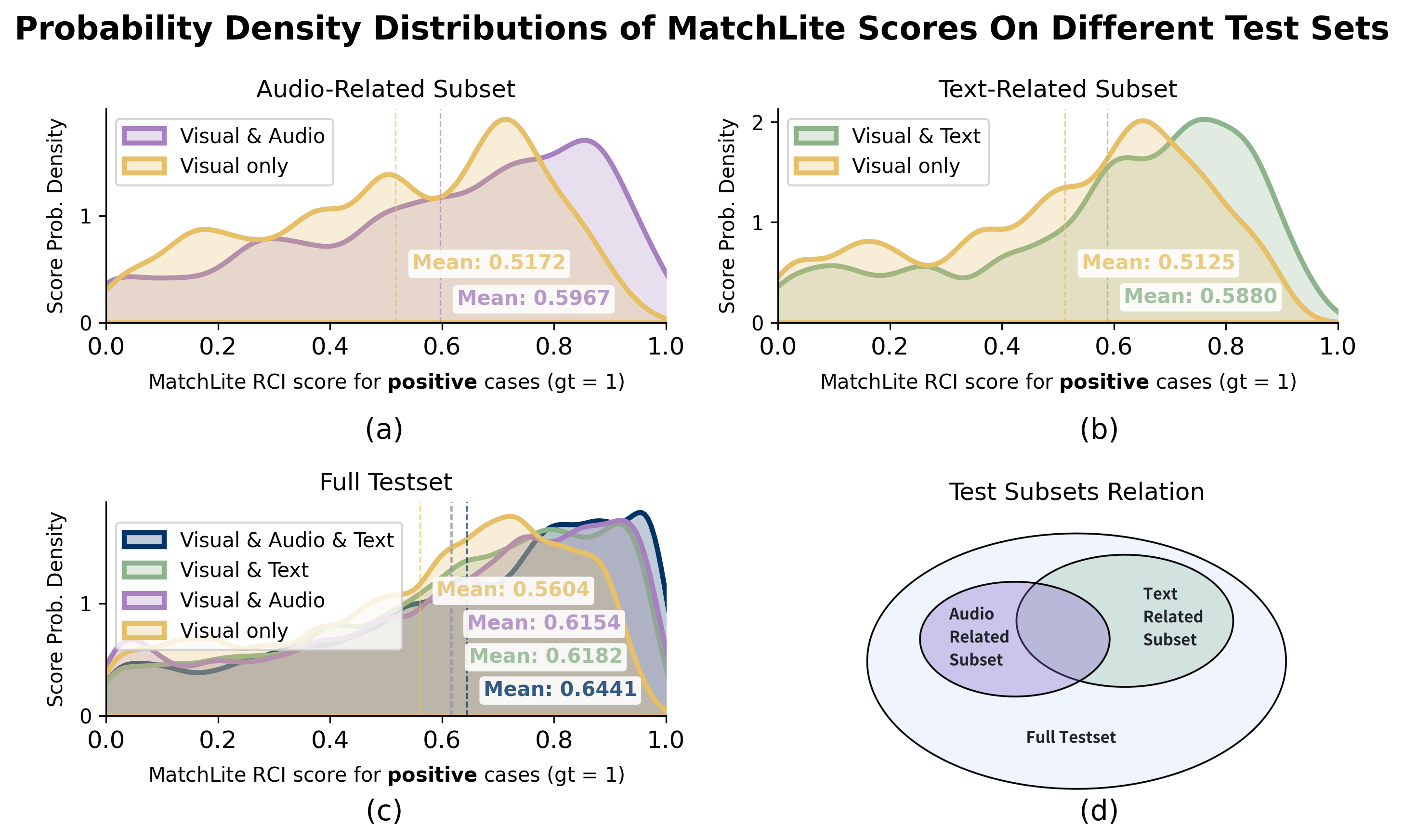}
  \caption{Different subsets are isolated from the full test set to observe the performance gains due to the addition of modalities. (a) For a subset whose labels are influenced by audio, adding audio improves performance. (b) For a subset whose labels are influenced by text, adding text improves performance. (c) For the full test set, combining audio and text yields the largest gains, while either modality alone still outperforms the visual-only model. (d) Relationships among the test subsets. }
  \Description{Adding audio modality correctly skews \textbf{MatchLite} RCI scores for reproduced content with auditory related reasons}
  \label{fig:audio_text_rci_scores_distribution_2x2_v5}
  \vspace{-14pt} 
\end{figure}

\subsubsection{Additional Modality Studies}

We investigate the benefits of incorporating additional modalities and compare the performance of \textbf{MatchLite} with 4 variants: 1) Baseline (Visual only \textbf{MatchLite}) 2) Visual + Audio \textbf{MatchLite} 3) Visual + Text \textbf{MatchLite} 4) Visual + Text + Audio \textbf{MatchLite}. As seen from Table ~\ref{tab:table1_ablations_lm_lite}, the inclusion of any of the additional modalities like audio and text will improve the performance of \textbf{MatchLite}, with the inclusion of the audio modality accounting for comparatively larger additional improvement (+3.39 AP/+3.26 F1) than the inclusion of the text modality (+0.83 AP/+0.94 F1). Finally, adding all 3 modalities will yield the best improvement of +3.68 AP/+3.56 F1. Similarly, for \textbf{MatchLM} we observe a performance boost when incorporating audio modality during training, accounting for an additional improvement of +3.99 AP/+2.27 F1.

To further analyze the impact of adding modalities, we extract targeted subsets from the main test set and visualize the score distributions for each model. As shown in Figure~\ref{fig:audio_text_rci_scores_distribution_2x2_v5}, adding audio shifts the score distribution in the positive cases audio-reason subset toward higher reproduction scores, and adding text produces a similar shift in the positive cases text-reason subset. Adding modalities also shifts the distribution for the positive cases in the full test set in the correct direction.
\subsubsection{Data Scaling}

Model capabilities are deeply shaped by training data characteristics \cite{mirzasoleiman2020coresets, qin2025infocoevolution}. We conduct data scaling experiments on \textbf{MatchLite} and \textbf{MatchLM} as shown in Table ~\ref{tab:table1_ablations_lm_lite}. \textbf{MatchLM} significantly outperforms \textbf{MatchLite} with only 1/3 of the training data, demonstrating that \textbf{MatchLM} is more sample efficient under data constrained conditions. We also leverage model checkpoints pre-trained on other in-house tasks into both \textbf{MatchLite} and \textbf{MatchLM}'s training and find an additional +2.21 AP/+0.37 F1 for \textbf{MatchLite} and an additional +1.22 AP/+0.59 F1 for \textbf{MatchLM}.

\subsubsection{Knowledge distillation}
Resource constraints present a real challenge when it comes to serving the full real-time traffic on our video sharing platform.
While \textbf{MatchLM} achieves the best performance, it is significantly more resource intensive than \textbf{MatchLite}, raising an important concern for cost-effectiveness during online deployment. 
Taking this into consideration, we instead apply knowledge distillation using \textbf{MatchLM} as the teacher model and the \textbf{MatchLite} as the student model. 

Concretely, we perform knowledge distillation from MatchLM to MatchLite via KL divergence on classification logits and cosine distance loss on embeddings, both with $\lambda = 1.5$, yielding a +9.71\% recall at P80 as seen in Table ~\ref{tab:model-comparison-lightweight-kd-mllm}. From our ablation studies, we find that the benefits of knowledge distillation plateau at $\lambda=1.5$. Increasing further to $\lambda=2$ yielded comparable results. Finally we selected the model trained with $\lambda = 1.5$ for deployment as it achieved the highest increase in AP (+1.86) and R@P80 (+9.71), with its F1 marginally trailing the $\lambda=2$ config by 0.2. Detailed ablations and analysis on varying $\mathcal{L}_{emb}$ and $\mathcal{L}_{logits}$ can be found in Appendix \ref{subsec:Effect of Knowledge Distillation}, along with visualisations of differences in the learned feature maps in MatchLite with and without distillation.

To verify that these gains are not simply attributable to model scaling, we also train a stronger non-MLLM \textbf{MatchLite+} variant in which the SwinTiny ($\sim$29M-parameter) visual encoder is replaced by SwinLarge ($\sim$243M parameters), using the same supervised data and no distillation (Table~\ref{tab:table1_ablations_lm_lite}, Distillation block). Although this $\sim$8$\times$ larger non-MLLM baseline reaches AP=82.73, F1=75.49, R@P80=69.81, our distilled \textbf{MatchLite+} (SwinTiny, w KD) attains comparable AP and stronger F1 (82.45/77.10) at a fraction of the parameter count, indicating that knowledge distillation from \textbf{MatchLM} provides gains beyond what is achievable from parameter scaling of the visual encoder alone.

A performance gap remains: the \textbf{MatchLM} achieves R@P80 of 75.92, with 9\% higher recall than the distilled \textbf{MatchLite}. We identify two possible reasons: (1) the feature backbones of the \textbf{MatchLite} are frozen during serving to support shared feature caching and minimize inference cost, which limits its ability to learn new patterns effectively; and (2) \textbf{MatchLite} has significantly fewer parameters than \textbf{MatchLM}, leading to lower representation capacity even with Knowledge Distillation.

\begin{figure}[h]
  \centering 
  \includegraphics[width=\columnwidth, keepaspectratio]{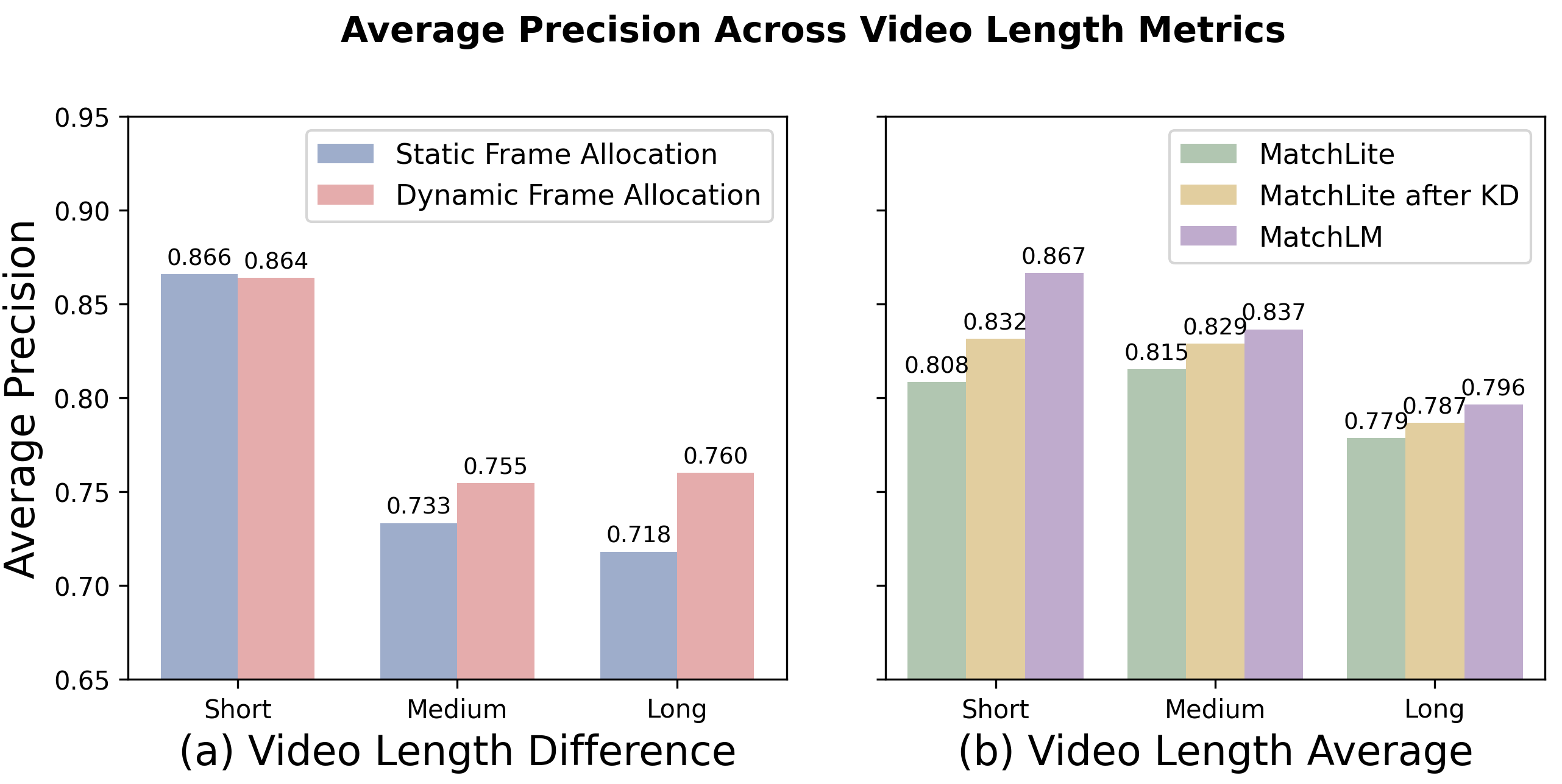}
  \caption{(a) Effectiveness of Dynamic Sampling versus Static Frame Allocation in MatchLM under different video length differences. Dynamic Sampling adapts better to mismatched video lengths, and its effects amplify as the length disparity between the video pair grows.
  (b) Performance of models across varying average video lengths. The capabilities of all models drop as the average length of video pairs increase, while KD proves effective across all video lengths.
  }
  \label{fig:f1_score_diff_dynamic_static}
  \vspace{-18pt} 
\end{figure}

\subsubsection{MatchLM Ablation Studies}

To study the contribution of individual components in \textbf{MatchLM}, we conduct several ablation experiments: (1) next-token prediction (NTP) formulated as a binary multiple-choice task (reproduced vs. not reproduced) versus last-token classification with a dedicated head; (2) early versus late audio fusion; (3) dynamic versus static frame allocation (with padding to 60 seconds); and (4) replacing the MLLM backbone (Qwen2.5-VL 3B) with LLaVA-OV 0.5B under the best-performing configuration. Results are summarized in Table~\ref{tab:mllm_ablation_studies}.

We adapt our task to the NTP paradigm by framing it as a multiple-choice question as a baseline for comparison with our last-token classification objective. Last-token classification consistently outperforms NTP, likely because it constrains the prediction space to two classes rather than the full vocabulary. More prompt details are provided in Appendix~\ref{subsec:MLLM NTP Details}.

We further compare early and late audio fusion strategies. For late fusion, we mean-pool audio embeddings and concatenate them with the MLLM’s last-token embedding before classification. For early fusion, the aggregated audio token is interleaved directly into the input token sequence. Implementation details are included in Appendix~\ref{sec:Audio Token Saliency Weighted Aggregation} and Appendix~\ref{subsec: MLLM Late Audio Fusion Details}. Early fusion outperforms late fusion, achieving gains of +1.48 AP and +0.88 F1. This improvement likely arises because early fusion enables richer cross-modal interactions among audio, visual, and textual tokens, whereas late fusion relies on a limited linear combination and cannot fully exploit the MLLM’s reasoning capacity.

Frame selection plays an important role in MLLM performance \cite{tang2025adaptive,zhu2025focus}. We experiment with dynamic frame-assignment strategies by allocating frames proportionally to each video's length. We find this outperforms static allocation in both AP (+0.16) and F1 (+0.67). As shown in Figure~\ref{fig:f1_score_diff_dynamic_static}(a), dynamic sampling is particularly beneficial when video pairs exhibit large length discrepancies, as it preserves more contextual information. Sampling details are provided in Appendix~\ref{sec:Dynamic Frame Allocation}. Applying the best configuration to the Qwen2.5-VL 3B backbone yields comparable but slightly lower performance than LLaVA-OV 0.5B (Table~\ref{tab:mllm_ablation_studies}).

\subsection{Cross-Domain Generalization}
\label{subsec:cross-domain-generalization}
We evaluate cross-domain generalization on two additional in-house datasets from production traffic. The \textbf{Short-Video (SV)} dataset contains 0.6M multimodal online short-video pairs, with approximately 20\% reproduced pairs and a 10\% subset held out for validation. The \textbf{Live-Streaming (L)} dataset contains 0.6M multimodal pairs of 20-second clips sampled from online live streams, with approximately 7\% reproduced pairs and a 10\% validation subset. 
The results are reported in Table~\ref{tab:cross-domain-results}. The distilled \textbf{MatchLM2Lite} improves over the visual-only baseline and substantially closes the gap to the teacher, demonstrating that the method is effective across distinct data distributions and tasks.

\begin{table*}[t]
  \centering
  \caption{Cross-domain generalization on internal Short-Video (SV) and Live-Streaming (L) datasets. \textbf{MatchLM2Lite} denotes the distilled \textbf{MatchLite} student trained with the \textbf{MatchLM} teacher; bolded numbers highlight the deployed lightweight model.}
  \vspace{-4pt}
  \label{tab:cross-domain-results}
  \begin{tabular}{lcrrrrrr}
    \toprule
    \multirow{2}{*}{Method} & \multirow{2}{*}{Modality} & \multicolumn{3}{c}{Short-Video (SV)} & \multicolumn{3}{c}{Live-Streaming (L)} \\
    \cmidrule(lr){3-5} \cmidrule(lr){6-8}
    & & AP & F1 & R@P80 & AP & F1 & R@P80 \\
    \midrule
    Baseline & V & 84.96 & 77.37 & 74.68 & 72.84 & 68.58 & 49.17 \\
    MatchLite & V+A+T & 87.57 & 80.11 & 78.73 & 77.52 & 69.60 & 60.07 \\
    MatchLM & V+A+T & 92.75 & 85.51 & 90.42 & 81.34 & 76.68 & 71.81 \\
    \textbf{MatchLM2Lite} & V+A+T & \textbf{88.70} & \textbf{81.26} & \textbf{81.83} & \textbf{79.50} & \textbf{74.78} & \textbf{68.76} \\
    \bottomrule
  \end{tabular}
  \vspace{-4pt}
\end{table*}

We further extend our \textbf{MatchLM} to a Hierarchical Reproduced Content Identification (H-RCI) task, which tries to determine whether a published video is reproduced from source content such as movies, TV shows, or dramas. We compare to a single-video input (mono-model) MLLM variation, and see that our paired, \textbf{MatchLM} can increase R@P80 from 58.72 to 86.5 on this task. More details can be found in Appendix \ref{subsec:H-RCI}.
\subsection{Public Benchmark Evaluation}
\label{subsec:public-benchmark-evaluation}
While our main experiments are conducted on an internal RCI dataset, we additionally evaluate the \textbf{MatchLM2Lite} framework on two widely used public video-copy detection benchmarks to verify external generalization: \textbf{VCSL} \cite{vcsl} and \textbf{VCDB} \cite{vcdb}. We compare against two competitive recent video-copy localization methods, TransVCL \cite{transvcl} and RTR \cite{rtr}.

\begin{table}[t]
  \centering
  \caption{Public-benchmark evaluation on VCSL and VCDB. We report video-level F-score (\%) for each model. \textbf{MatchLM2Lite} denotes the distilled lightweight student.}
  \vspace{-4pt}
  \label{tab:public-benchmark}
  \setlength{\tabcolsep}{4pt}
  \begin{tabular}{@{}lrr@{}}
    \toprule
    Model & VCSL F-score & VCDB F-score \\
    \midrule
    TransVCL \cite{transvcl} & 90.06 & 87.94 \\
    RTR \cite{rtr}           & 96.46 & 94.12 \\
    MatchLite                & 95.53 & 93.27 \\
    MatchLM                  & 99.04 & 99.32 \\
    \textbf{MatchLM2Lite}    & \textbf{98.74} & \textbf{97.23} \\
    \bottomrule
  \end{tabular}
  \vspace{-4pt}
\end{table}

As shown in Table~\ref{tab:public-benchmark}, the teacher \textbf{MatchLM} achieves near saturation performance on both benchmarks (99.04 on VCSL, +2.58 over RTR; 99.32 on VCDB, +5.20 over RTR). The distilled \textbf{MatchLM2Lite} retains most of the teacher's performance after distillation, surpassing RTR by +2.28 on VCSL and +3.11 on VCDB while operating at the much smaller \textbf{MatchLite} parameter and serving-cost budget. These results demonstrate that the multimodal pairwise-matching knowledge transferred from \textbf{MatchLM} to \textbf{MatchLite} generalizes beyond the internal production dataset to standard public video-copy benchmarks. We note that VCSL and VCDB focus narrowly on copy detection rather than policy-grounded multimodal reasoning, so the internal RCI task remains the more complex and policy-dependent task.

\section{Online Serving and Experiments}

\subsection{Online Serving}
\label{sec:Online serving}
\textbf{MatchLite} requires only 0.86 TFLOPs per inference, compared to 25.4 TFLOPs for \textbf{MatchLM}, yielding a 35$\times$ reduction in serving-time computational cost on an NVIDIA A10 24GB GPU. Leveraging knowledge distillation, we preserve \textbf{MatchLite}'s efficiency while achieving a 9.71\% improvement in R@P80 under the same computational budget.

We integrate the best \textbf{MatchLite} model into our online systems for RCI detection. When a video is published, an upstream retrieval module first selects the top-1 candidate from a vector database based on visual similarity. The original and retrieved videos are then passed to \textbf{MatchLite}, which produces an RCI score used by the recommendation system to prioritize original content over reproduced ones. All newly published videos on the platform are scored in real time, and those with high RCI scores are deboosted (less likely to be recommended to users). In our production deployment, retrieval latency is approximately 12s, \textbf{MatchLite} inference latency is about 2.8s, and the end-to-end pipeline latency remains below 30s. The overall system stays stable under an average load of 2.8k queries per second (QPS), with peak QPS exceeding 3.5k.

\subsection{Online Experiments}
We conducted a two-week online A/B experiment on our short video platform, allocating 10\% of the overall online video traffic across 2 groups. Users were randomly assigned to either a control or treatment group. RCI scores used for moderation in the control group were determined by a visual-only online baseline model, while the treatment group deployed the distilled \textbf{MatchLite} to identify reproduced content. We evaluated the impact on two key business metrics: stay duration (the average time users spend on the platform, measuring user engagement) and reproduced video views (the number of viewed videos that are reproduced). \textbf{MatchLite} reduced the reproduced-video view rate by 2.5\% without significantly affecting user stay duration. Throughout the A/B experiment, we monitored core user metrics and platform health indicators to assess both effectiveness and potential side effects. After post-experiment review, we deployed \textbf{MatchLite} to serve full online traffic.

\section{Conclusion and Future Work}
In this work, we propose a \textbf{MatchLM2Lite} framework for reproduced content identification (RCI) in our short video platform. Our approach utilizes a powerful MLLM teacher, \textbf{MatchLM}, to learn strong modality alignment and rich semantic representations aligned to the RCI policy. We then distill its knowledge to \textbf{MatchLite} and achieve low-latency and high-throughput inference in production scenarios while maintaing strong performance. This framework demonstrates a practical and scalable solution for
large-scale industry deployment, while also opening up new directions for multimodal video copy detection using MLLM. 

For future work, we will further explore the capabilities of MLLMs by aligning to user feedback through continual learning stategies. We also plan to extend our approach to understanding of longer video sequences through token merging or compression strategies. Beyond classification of reproduced content, we aim to leverage MLLMs for temporal grounding of copied video segments, enabling more fine-grained and explainable content copy detection.

\newpage
\bibliographystyle{ACM-Reference-Format}
\balance
\bibliography{references}


\appendix
\section*{Appendix}
\section{Architecture Details}
\label{sec:Architecture details}

\subsection{MatchLite Details}
\label{subsec:MatchLite Details}
\paragraph{Implementation Details}

For each video, we sample visual/audio binaries at 1 FPS and extract user-generated text. Frames are dynamically sampled up to the maximum supported length, zero-padded when shorter, resized to 224 x 224, and encoded by an in-house pretrained Swin-T encoder into 128-dimensional visual embeddings. Audio is encoded by an in-house Whisper-small encoder into 1500 x 768 embeddings and mean-pooled to 1 x 768 for serving efficiency. For text, we concatenate title and sticker text, then use multilingual Sentence-BERT MiniLM \cite{wang2020minilm} to obtain 64 x 384 embeddings with 64-token truncation. The model uses an 8-head BiXT module with hidden size 64, a 3-layer Conv2D visual enhancement module with BatchNorm/ReLU/residual connections, and a ResNet-34 decision module whose first convolution is resized to match the feature-channel count.

\subsection{MatchLM Details}
\label{subsec:MatchLM Details}

\paragraph{Dynamic Frame Allocation}
\label{sec:Dynamic Frame Allocation}
We allocate MatchLM frames according to the video-length ratio, while preserving a minimum allocation for the shorter video when possible. Frames are uniformly sampled under a fixed pair-level frame budget and prefixed with "Video 1:" and "Video 2:" to distinguish the two inputs.
\paragraph{Audio Token Saliency Weighted Aggregation}
\label{sec:Audio Token Saliency Weighted Aggregation}
Audio is interleaved into MatchLM with an "Audio:" prefix for each video. A Whisper-small encoder extracts 1500 x 768 audio embeddings; a learned linear layer produces token saliency scores, which are softmaxed to form a weighted audio token. This token is projected to the text-token dimension before being fed into the LLM.

\section{Ablation details}
\label{sec:Ablation details}

\subsection{Reframing Reproduced Content Identification as Next Token Prediction}
\label{subsec:MLLM NTP Details}

To adapt RCI to Next Token Prediction (NTP), we frame it as a multiple-choice task using the prompt "Choose the correct option from the given options: A) Normal B) Reproduced. Answer:". We train the model to predict A for normal and B for reproduced content. During validation, we softmax the last-token logits for "A" and "B" as the two class probabilities, providing a direct comparison to our last-token classification objective.

\subsection{Late Audio Fusion}
\label{subsec: MLLM Late Audio Fusion Details}
For late audio fusion, we concatenate the mean-pooled 1x768 audio embeddings from both videos, pass them through a linear layer, concatenate the result with the MLLM last-token embedding, and feed it to the classification head. The model is trained with the same cross-entropy loss as early fusion.

\subsection{Effect of Knowledge Distillation}
\label{subsec:Effect of Knowledge Distillation}

\begin{figure}[h]
  \centering
  
  \begin{minipage}[t]{0.46\linewidth} 
    \centering
    \underline{\textbf{Example 1}}
  \end{minipage}
  \hspace{0.04\linewidth} 
  \begin{minipage}[t]{0.46\linewidth} 
    \centering
    \underline{\textbf{Example 2}}
  \end{minipage}
  
  \vspace{0.2em} 
  
  \begin{minipage}[t]{0.23\linewidth}
    \centering
    \textbf{After KD}
  \end{minipage}
  \hfill
  \begin{minipage}[t]{0.23\linewidth}
    \centering
    \textbf{Before KD}
  \end{minipage}
  \hspace{0.04\linewidth} 
  \begin{minipage}[t]{0.23\linewidth}
    \centering
    \textbf{After KD}
  \end{minipage}
  \hfill
  \begin{minipage}[t]{0.23\linewidth}
    \centering
    \textbf{Before KD}
  \end{minipage}
  
  \vspace{0.3em} 
  
  \begin{minipage}[t]{0.23\linewidth}
    \centering
    \includegraphics[width=\linewidth]{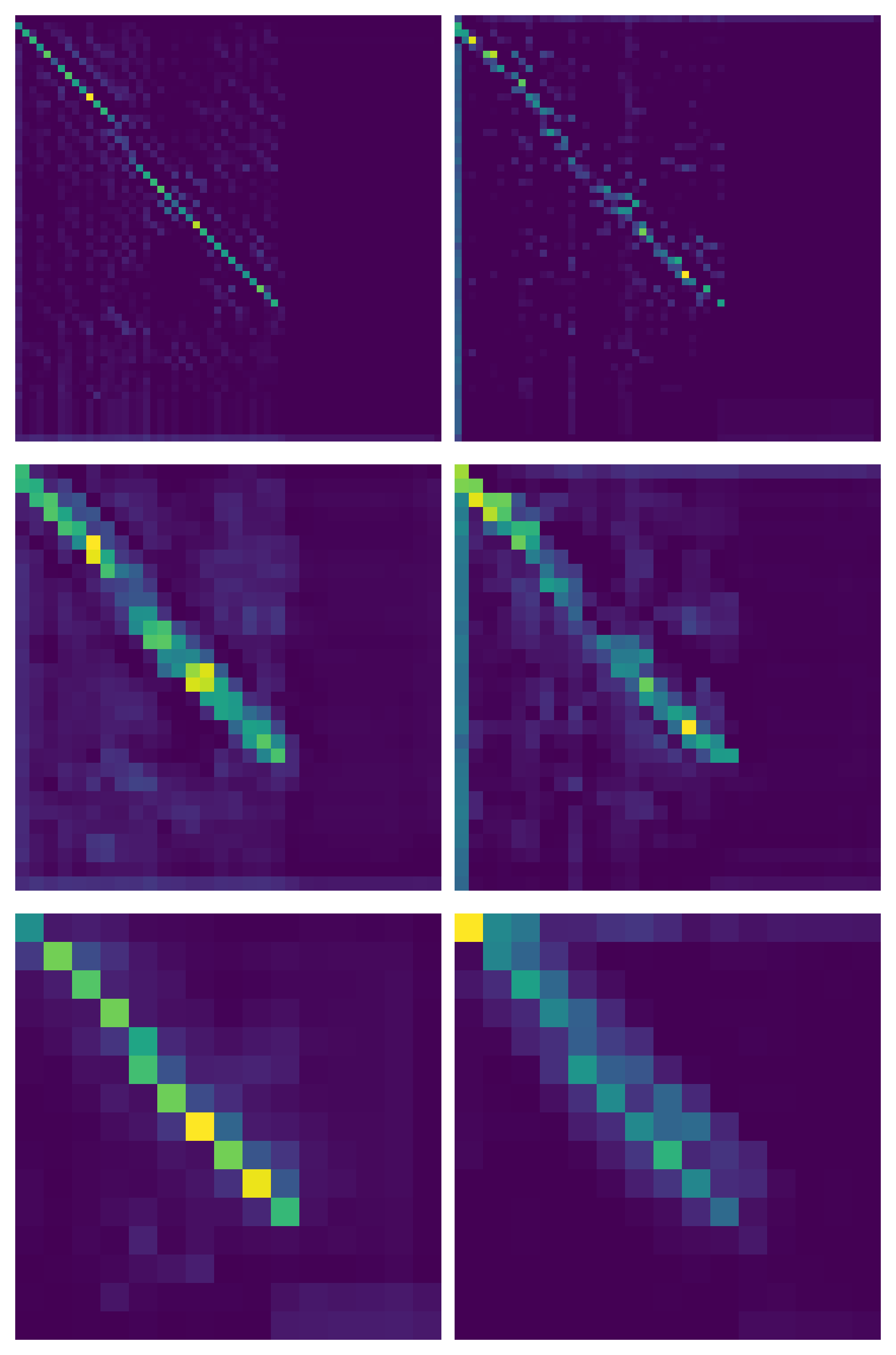}
    \label{fig:ex1_after}
  \end{minipage}
  \hfill
  \begin{minipage}[t]{0.23\linewidth}
    \centering
    \includegraphics[width=\linewidth]{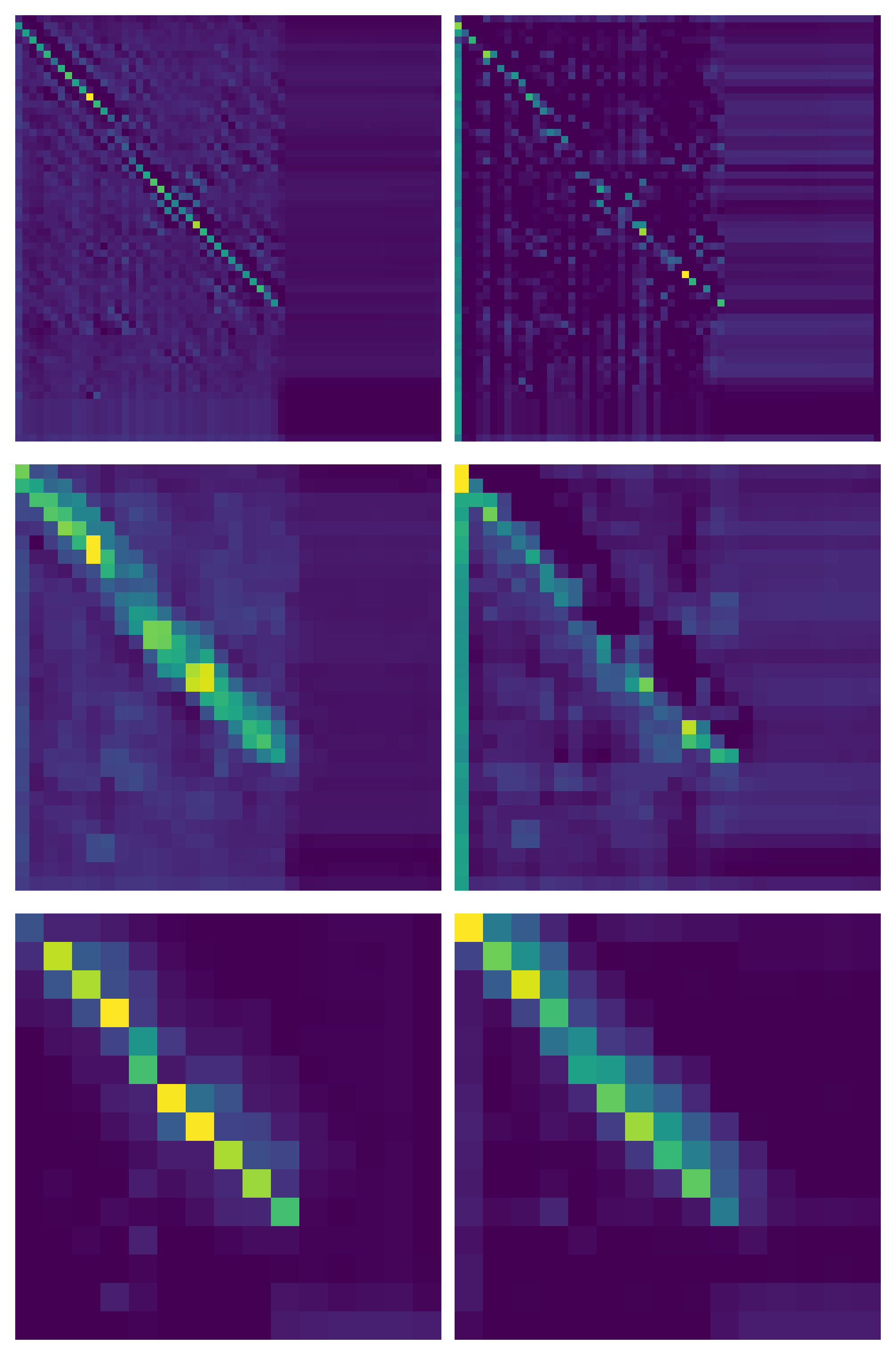}
    \label{fig:ex1_no}
  \end{minipage}
  \hspace{0.04\linewidth} 
  \begin{minipage}[t]{0.23\linewidth}
    \centering
    \includegraphics[width=\linewidth]{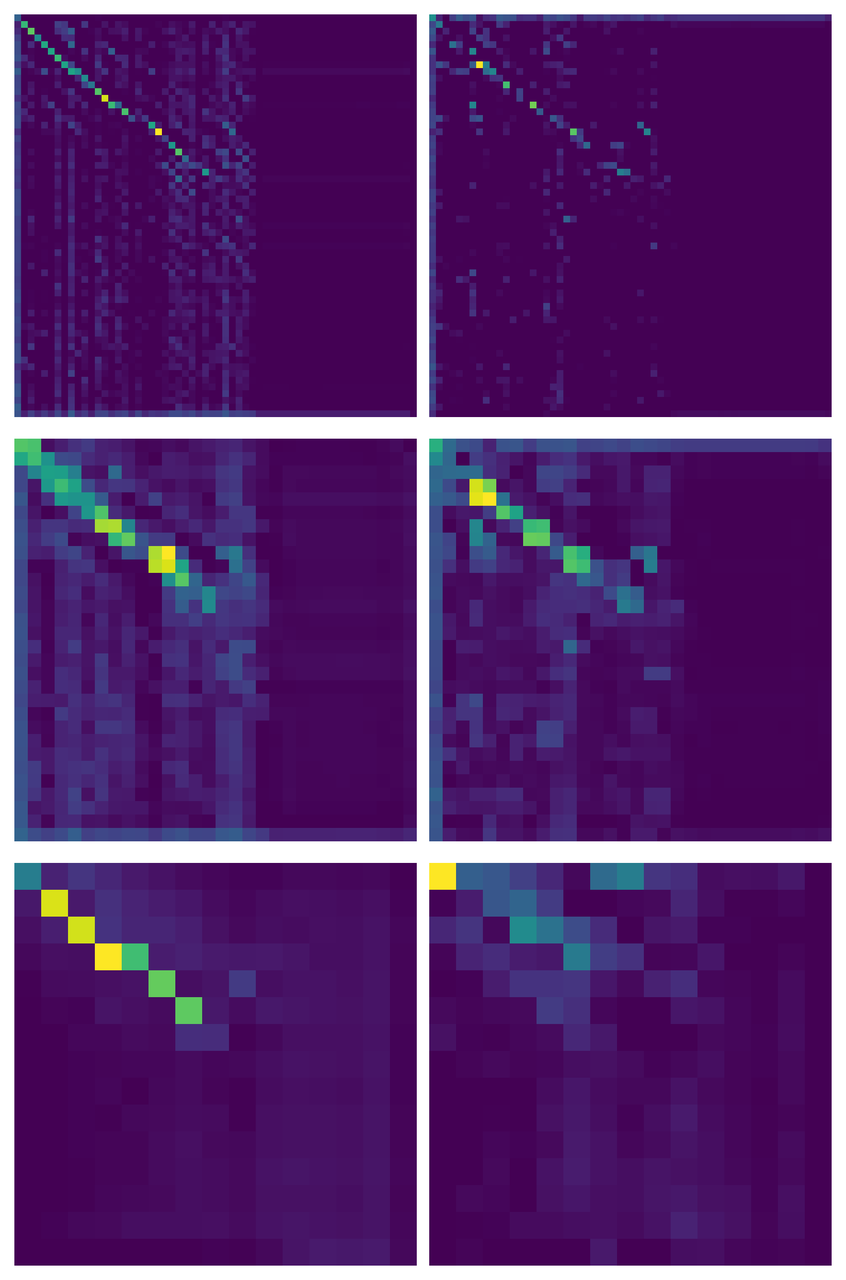}
    \label{fig:ex2_after}
  \end{minipage}
  \hfill
  \begin{minipage}[t]{0.23\linewidth}
    \centering
    \includegraphics[width=\linewidth]{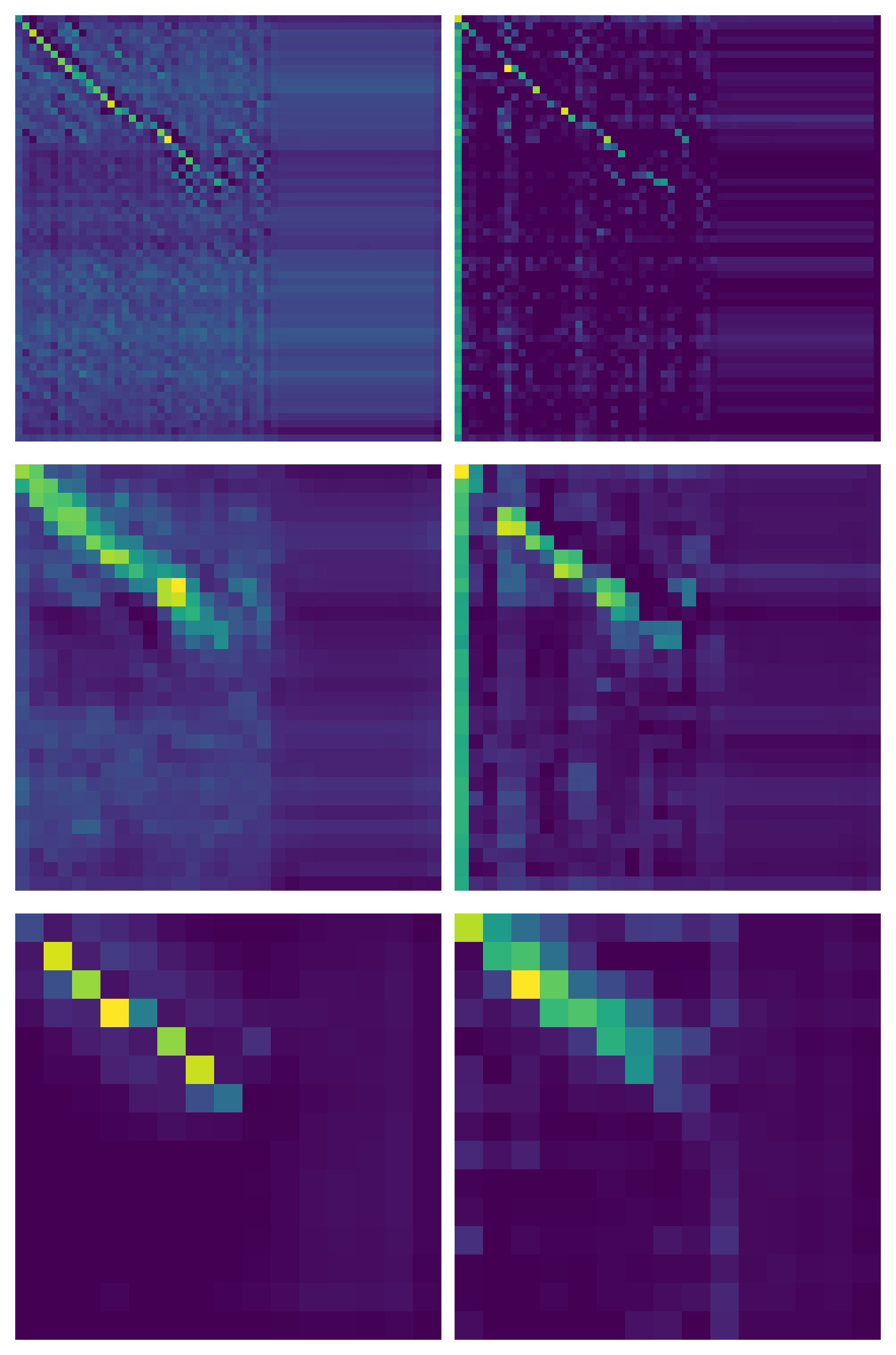}
    \label{fig:ex2_no}
  \end{minipage}
  \vspace{-16pt}
  \caption{Feature maps after Knowledge Distillation qualitatively appear to be slightly more distinct, ignoring background noise features}
  \Description{Small model feature maps.}
  \label{fig:feat_comparison}
  \vspace{-8pt}
\end{figure}

Table~\ref{tab:model-comparison-lightweight-kd-mllm-logit-weight-ablation} reports KD ablations. KL loss alone at $\mathcal{L}_{logits}=1.5$ improves F1/AP/R@P80 to 77.30/82.26/65.34, while embedding loss alone at $\mathcal{L}_{emb}=1.5$ reaches 76.03/81.65/62.66. Increasing either loss while holding the other fixed at 1.5 generally improves performance, with the best setting at $\mathcal{L}_{emb}=1.5$ and $\mathcal{L}_{logits}=1.5$.

\begin{table}
  \centering
  \caption{Knowledge Distillation ablations on MatchLite+.}
  \label{tab:model-comparison-lightweight-kd-mllm-logit-weight-ablation}
  \vspace{-6pt}
  \setlength{\tabcolsep}{3pt}
  \begin{tabular}{lccrrr}
    \toprule
    Setting & $\mathcal{L}_{\text{logits}}$ & $\mathcal{L}_{\text{emb}}$ & AP & F1 & R@P80\\
    \midrule
    Baseline & -- & -- & 80.59 & 74.48 & 57.21\\
    Emb. only & -- & 1.5 & 81.65 & 76.03 & 62.66\\
    Logit only & 1.5 & -- & 82.26 & 77.30 & 65.34\\
    \midrule
    Vary logit & 0.5 & 1.5 & 82.51 & 76.51 & 65.57\\
    Vary logit & 1.0 & 1.5 & 82.37 & 76.54 & 65.02\\
    Vary logit & 1.5 & 1.5 & 82.45 & 77.10 & 66.92\\
    Vary emb. & 1.5 & 0.5 & 81.90 & 76.25 & 66.59\\
    Vary emb. & 1.5 & 1.0 & 82.46 & 76.83 & 66.24\\
    \bottomrule
  \end{tabular}
\end{table}

Qualitatively, Figure~\ref{fig:feat_comparison} shows clearer Compact Decision feature maps for reproduced pairs after KD, suggesting that \textbf{MatchLM} guidance helps \textbf{MatchLite} learn more representative pairwise patterns. Across average video lengths, KD improves \textbf{MatchLite} for short, medium, and long videos, as shown in Figure~\ref{fig:f1_score_diff_dynamic_static}(b).

\section{Hierarchical Reproduced Content Identification: H-RCI}
\label{subsec:H-RCI}
We define Hierarchical Reproduced Content Identification (H-RCI) as determining whether a published video is reproduced from source content using both single-video understanding and paired-video comparison. To address H-RCI, we extend the original \textbf{MatchLM} into a unified model that handles both paired and single-video inputs.
H-RCI is a supplementary application showing that \textbf{MatchLM} extends beyond paired-video RCI. The joint model accepts single- and paired-video inputs, combining source-content reproduction detection (e.g., movies, TV shows, dramas) with paired comparisons. This task is orthogonal to the production RCI pipeline and distillation framework in the main paper. 

As shown in Table~\ref{tab:Joint model on H-RCI task}, mono MLLM reaches R@P80 of 58.72. Combining mono and paired MLLMs improves R@P80 to 81.02, and the joint model further raises it to 86.5 by using paired-video input as a pre-filter for the video-understanding branch.
\begin{table}[h]
  \centering
  \setlength{\tabcolsep}{3pt}
  \caption{Joint model on H-RCI task}
  \label{tab:Joint model on H-RCI task}
  \begin{tabular}{@{}lrrr@{}}
    \toprule
    Model&AP&F1&R@P80\\
    \midrule
    Mono Model & 72.93 & 67.73 & 58.72\\
    Mono + Paired model & - & 80.5 & 81.02\\
    Joint Model & 89.2 & 83.12 & 86.5 \\
  \bottomrule
\end{tabular}
\end{table}

\section{Training Details}
\label{sec:Training Details}

\subsection{MatchLite Model}
\label{subsubsec:MatchLite Model}

\textbf{MatchLite} has 263M parameters including frozen feature backbones. It was trained for 8 epochs on 16 NVIDIA A100 80GB GPUs with batch size 16 per GPU, AdamW, cosine annealing, and learning rate 1e-4. The feature extractors are frozen for feature caching and faster deployment; the remaining layers are trained.

\subsection{MatchLM Model}
\label{subsubsec:MatchLM Model}

To manage memory, \textbf{MatchLM} uses LLaVA-OV 0.5B, dynamic frame allocation, and early audio fusion with learned saliency pooling. The 1B-parameter model was trained for 8 epochs on 48 A100 GPUs with batch size 1 per GPU, AdamW, linear annealing, and learning rate 2e-5. Training uses ZeRO Stage 2 \cite{rasley2020deepspeed} and LoRA with rank 32, alpha 64, and dropout 0.05.

\section{Case Studies}
\label{sec:Case Studies}
We present four anonymized case studies, omitting frames for privacy and report scores from each \textbf{MatchLite} variant and the \textbf{MatchLM} teacher.

\paragraph{Case 1: Visual-only (V)}
\label{subsec:case-v}The query video shows a person surfing at the beach, riding waves, smiling on a surfboard, and walking out of the ocean; its audio is upbeat pop music and it has no visible text. The candidate uses the same frame sequence but replaces the audio with different upbeat rock music and still has no text. Because the visual stream is frame-identical, V alone saturates, and audio/text only add small calibration gains.

\paragraph{Case 2: Visual + Audio (V+A)}
\label{subsec:case-va}The query video is a montage of a pop singer performing across multiple stage setups and outfits, paired with a popular K-pop song and no visible text. The candidate shows the same artist and stage production, but with different camera angles and shot selections; its audio is identical to the query and it also has no text. The visual stream is therefore ambiguous, while the shared song provides the decisive V+A signal.

\paragraph{Case 3: Visual + Text (V+T)}
\label{subsec:case-vt}
The query is a podcast-style discussion between two speakers, with dialogue audio and the full conversation rendered as bold white subtitles. The candidate contains the same podcast segment, but the video is mirrored, filtered with an old-film effect, and zoomed in; its audio keeps the same dialogue with a faint instrumental overlay, and the same subtitles appear in small green text at the bottom. The visual transformation and audio overlay weaken V and V+A, while matching subtitles provide the largest gain.

\paragraph{Case 4: Visual + Audio + Text (V+A+T)}
\label{subsec:case-vat} The query is a speech or sermon compilation with Indonesian quote overlays, a static background, full speech audio, and text quotes from the speech. The candidate contains partially overlapping speech excerpts by the same speaker, a different static background, decorative quote overlays, the speaker's voice mixed with emotional music, and a mix of matching and non-matching quotes. Because each modality is only partially reproduced, no single modality is decisive; joint V+A+T fusion is required.

\begin{table}[h]
  \centering
  
  \caption{Case-study scores across modality configurations.}
  \label{tab:case-study-scores}
  \vspace{-6pt}
  \small
  \setlength{\tabcolsep}{3pt}
  \begin{tabular}{@{}lrrrrr@{}}
    \toprule
    Case & V & V+A & V+T & V+A+T & MLLM \\
    \midrule
    1: V & 0.9339 & 0.9716 & 0.9668 & 0.9784 & 0.9961 \\
    2: V+A & 0.4892 & 0.7997 & 0.6342 & 0.8782 & 0.9023 \\
    3: V+T & 0.6138 & 0.8077 & 0.9232 & 0.9551 & 0.9726 \\
    4: V+A+T & 0.2044 & 0.5678 & 0.6535 & 0.7834 & 0.8750 \\
    \bottomrule
  \end{tabular}
  \vspace{-8pt}
\end{table}

\end{document}